\documentclass[
12pt]{article}
\usepackage{color}
\usepackage{graphicx}
 \usepackage{amsmath,amssymb,array,calc,rotating,epsfig,psfrag}  
\usepackage[nosort]{cite}

\definecolor{darkgreen}{rgb}{0,0.55,0}

\newcommand{\bea}{\begin{eqnarray}}
\newcommand{\eea}{\end{eqnarray}}
\newcommand{\be}{\begin{equation}}
\newcommand{\ee}{\end{equation}}

\def\p{\phi}

\def\revise#1       {\raisebox{-0em}{\rule{3pt}{1em}}%
                     \marginpar{\raisebox{.5em}{\vrule width3pt\  
                     \vrule width0pt height 0pt depth0.5em  
                     \hbox to 0cm{\hspace{0cm}{%
                     \parbox[t]{4em}{\raggedright\footnotesize{#1}}}\hss}}}}

\def\sqr#1#2{{\vcenter{\vbox{\hrule height.#2pt    
 \hbox{\vrule width.#2pt height#1pt \kern#1pt  
 \vrule width.#2pt}\hrule height.#2pt}}}}  
\def\square{%
  \mathop{\mathchoice{\sqr{12}{15}}{\sqr{9}{12}}{\sqr{6.3}{9}}{\sqr{4.5}{9}}}}  
 
\catcode`\@=12

\begin{document}

\makeatletter \@addtoreset{equation}{section} \makeatother  
\renewcommand{\theequation}{\thesection.\arabic{equation}}  

\renewcommand\baselinestretch{1.25}
\setlength{\paperheight}{11in}
\setlength{\paperwidth}{9.5in}
\setlength{\textwidth}{\paperwidth-2.4in}     \hoffset= -.3in   
\setlength{\textheight}{\paperheight-2.4in}   \topmargin= -.6in 

\begin{center}

{\large \bf Analytic Scattering Amplitudes for QCD}

\vskip .7cm

{\large Diana Vaman${}^1$ and York-Peng  Yao${}^2$}

\vskip .7cm

${}^1$ Department of Physics, University of Virginia,\\
Charlottesville, VA, 22904, USA\\
E-mail: dv3h@virginia.edu


${}^2$ Department of Physics, University of Michigan,\\
Ann Arbor, MI, 48109, USA\\
E-mail: yyao@umich.edu

\vskip .7cm

\begin{abstract}
By analytically continuing QCD scattering amplitudes through specific
complexified momenta, one can study and learn about the nature and the 
consequences of factorization and unitarity.  In some cases, when coupled 
with the largest time equation and gauge invariance requirement, this approach 
leads to recursion relations, which greatly simplify the construction of multi-gluon 
scattering amplitudes. The setting for this discussion is in the 
space-cone gauge.
\end{abstract}

\end{center}

\section{Introduction}

The LHC will be turned on soon.  Excluding serendipitous events, it
seems that signals will be seen only after complicated backgrounds
have been properly subtracted out.  Therefore, one must have a good
account of the multi-particle processes, particularly those induced 
by QCD.  Also, one should know the proper energy scale in a calculation
to ensure stability relative to higher order effects.  In other words, loops 
are also important, besides tree level results.  

There has been much progress in perturbative evaluations 
\cite{Bern:1994cg}, especially 
 in the past few years 
\cite{Witten:2003nn, Cachazo:2004kj,Britto:2004ap,Britto:2005fq}.  
One would even venture to say that there is 
 a new technology, which is applicable to all field theories.  We 
 shall confine our attention to QCD here.

If one is to follow the usual Feynman rules and diagrams to calculate
a multi-gluon process, one will find that the algebra becomes 
horrendous very fast.  For an $n$-gluon process at the tree level, if we 
just examine the three-point vertices, there are $n-2$ of them and each 
one has six terms which depend on some momenta, not to mention
the internal symmetry coupling.  Then one has to permute these $n$
legs over the vertices.

As we all know, a massless particle in four spacetime 
dimensions has at most two degrees
of freedom, but a manifestly covariant formulation requires four components 
for a vector field.  Therefore, there are tremendous amount of cancellations
in the intermediate stage of a calculation to yield some simple looking
final answer.  The process somehow knows that the unphysical degrees of 
freedom should not be there and tries its best to expel them.  Lots of efforts 
were wasted in the old ways.   
It will help if one eliminates all these unwanted degrees of 
freedom at an early stage in some way.

The new developments in non-Abelian gauge field calculations on the whole 
pursue two different paths, the ideas of which are not entirely new, 
but the executions are much improved:

(1)  Using a physical gauge, such that there are explicitly only two components
for each internal symmetry index.

(2) Using an extended dispersive technique, such that an $n$-point amplitude 
will be constructed from lower point on-shell physical amplitudes.

It turns out that these two methods can be made to complement each other 
and give rise to recursion relations for all the tree and some of the one-loop 
amplitudes \cite{Bern:2005hs}.  
For the other one-loop amplitudes with more complicated
helicity composition, they are very much like the 
dispersion method using Cutkosky rules, but of course with much better 
handle and insight.

You must appreciate the possibility of having recursion relations, because
one can recycle whatever hard work one has already put in to build up
more complicated processes, rather than to start from the scratch all 
over again.  This is possible, much to the credit of analytic continuation into complex 
momenta \cite{Britto:2005fq}.

\section{Spinors, Twistors and Complex Momenta}

For a particle with zero mass, we can use two component 
spinors or twistors representation
\begin{equation}
P^{\dot a b}=(\tilde \sigma \cdot P)^{\dot a b} =|p]^{\dot a}<p|^b
,
\end{equation}
and
\begin{equation}
P_{ b \dot a}=( \sigma \cdot P)_{ b \dot a} =|p>_b[p|_{\dot a}
\end{equation}
where $ \sigma^\mu =(-I, \vec \sigma)$ and $ \tilde \sigma^\mu =(-I, -\vec \sigma)$.
We use them to form scalar products of spinors
\begin{equation}
<p_ip_j>=<p_i|^b|p_j>_b=-<p_jp_i>,
\end{equation}
and 
\begin{equation}
[p_jp_i]=[p_j|_{\dot a}|p_i]^{\dot a}=-[p_ip_j],
\end{equation}
from which the scalar product of two vectors is
\begin{equation}
-2P_i\cdot P_j=<p_ip_j>[p_jp_i].
\end{equation}

Also, we use them to build polarization vectors for gauge
particles of momentum $K_i$\cite{ab}
\begin{equation}
\epsilon ^{h=+ }(Q_i,K_i)^\mu= {<q_i|\sigma ^\mu
|k_i]\over \sqrt 2 <q_ik_i>}, \ 
\epsilon ^{h=- }(Q_i,K_i)^\mu= {[q_i|\tilde \sigma ^\mu
|k_i>\over \sqrt 2 [q_ik_i]}
\end{equation}
in which $Q_i$ is a reference momentum, which can be individually
assigned for each $K_i$.  Changing $Q_i$ is a change of gauge.

For real momenta, we have
\begin{equation}
[p_ip_j]=<p_jp_i>^\star ,
\end{equation}
which is a result we don't like, if we want to perform on-shell calculation. 
Let us consider forming an amplitude for three on-shell gluons  
\begin{equation}
P_1+P_2+P_3=0.
\end{equation}
Then for real momenta
\begin{equation}
0=P_1^2=(P_2+P_3)^2=2P_2\cdot P_3 =-|<p_2p_3>|^2, \ \ etc.
\end{equation}
which means both
\begin{equation}
<p_i p_j>=0, \ \  {\rm and} \ \ [p_i p_j]=0, \ \ \ i,j=1,2,3.
\end{equation}
This makes it impossible to define an on-shell tree level 
three-point  gluon amplitude, 
which is the least demand to start a program.

On the other hand, for complex P's, $[p_jp_i]$ is no longer the 
complex conjugate of $<p_i p_j>$ and for appropriate helicity 
arrangements the zero mass conditions can be satisfied by either 
\begin{equation}
<p_i p_j>=0,
\end{equation}
then e.g. \cite{parketaylor} 
\begin{equation}
A(P_1^+,P_2^+,P_3^-)=-i{[p_1p_2]^4\over [p_1p_2][p_2p_3][p_3 p_1]},
\end{equation}
or
\begin{equation}[p_ip_j]=0,
\end{equation}
then e.g. \cite{parketaylor}
\begin{equation}
A(P_1^-,P_2^-,P_3^+)=i{<p_1p_2>^4\over <p_1p_2><p_2p_3><p_3 p_1>}.
\end{equation}

\section{Space-Cone Gauge}

There are many different physical gauges to get rid of the unphysical degrees
of freedom, but the one which is best for our purpose is the space-cone 
gauge.  This is because for the specific analytic continuation into complex
momenta we use to 
arrive at recursion relations,  we shall find that
the vertices in this gauge are untouched.  Let us be reminded that our
aim is to factorize each term in an amplitude, which has both a numerator
and a denominator,  into something simpler, already known or done.  
If we don't have to touch the numerator in our manipulation to 
accomplish this, it will be just that much easier.  In other words, we 
shall find that for this gauge the factorization is like what is needed
in a scalar theory, where we shall be massaging products of propagators
into something we can identify with a lower point on-shell amplitude. 

Although we are free to have one reference vector $(Q_i)$ for each emitted
gluon, we shall use only two reference spinors for all gluons
\begin{equation}
|+>, \ \ \ \ \ [-|
\end{equation}
and normalize them to
\begin{equation}
<+->=[-+]=1.
\end{equation}

Any massless four vector can be decomposed according to
\begin{equation}
P=p^+|->[-|\ + \ p^-|+>[+|\ +\ p|->[+| \ + \ \bar p|+>[-|,
\end{equation}
and a gluon of momentum K has polarization vectors
\begin{equation}
\epsilon ^+(K)={[-k]\over <+k>}, \ \ \ \epsilon ^-(K)={<+k>\over [-k]}.
\end{equation}
They satisfy 
\begin{equation}
\epsilon^+(K)\epsilon^-(K)=1
\end{equation}
which makes polarization sums very simple.

The space-cone gauge \cite{Chalmers:1998jb} is defined by the condition 
\begin{equation}
N\cdot A=0 \ \ \ \ or \ \ \ a=0
\end{equation}
for each color index of the gauge field A.  Here
\begin{equation}
N=|+>[-|
\end{equation}
is also a light-like vector.  There is a constraint among the
equations of motion, which can be used to express $\bar a$
in terms of $a^\pm$, and the resulting Lagrangian is 
\begin{eqnarray}
L&=&Tr \bigg( {1\over 2} a^+\partial _\mu\partial^\mu a^-
-i\big({\partial ^-\over \partial}a^+\big)[a^+, \ \partial a^-]
\nonumber\\
&-&i\big({\partial ^+\over \partial}a^-\big)[a^-, \ \partial a^+]
+[a^+, \ \partial a^-]{1\over \partial^2}[a^-, \ \partial a^+]\bigg)
\end{eqnarray}
What is noteworthy is that in the interaction part of $L$ we do not 
have the derivative component $\bar \partial $, which is very important for later 
discussion when we perform analytic continuation by shifting
momenta, or derivatives.  We shall find that only $\bar \partial$
will be affected, but this does not appear in the vertices, which means
that the interaction will be unchanged.  However, all components 
$\partial, \ \bar \partial, \  \partial ^\pm$ appear in the Klein-Gordon 
operator $\partial _\mu \partial^\mu$, and therefore propagators will change 
when we do analytic continuation.  It is as if we have a two component scalar 
field theory.

The analysis is further simplified by color ordering. 

\section{The Largest Time Equation and Analytic Continuation}

The causal nature of quantum field theory allows one to decompose
a propagator into a positive frequency part and a negative frequency 
part.  From this, identities for products of propagators and products
in which some propagators are replaced by positive or negative 
parts can be written down.  One easy way to arrive at them is to
observe that if a system is driven from $t=-\infty $ to $t=+\infty $
by some external currents and then back to $t=-\infty$, the generating
functional must be just unity.  By equating the coefficients of various 
powers of external currents, one can obtain sets of identities. The
physical outcome is that for every Feynman diagram in a scattering process, 
one can draw boundaries with inflowing energy lines on one side
and outflowing energy lines on the other.  The largest time equation by 
Veltman \cite{Veltman:1963th}, (which is closely associated with the closed time path cycle
of Schwinger \cite{schwinger}), is to pick two out of possibly many space-time points 
in a diagram and time order them.  This will relate a product of propagators 
with cut lines.  The causal ordering is enforced by a parameter z, which is an
integration variable
\begin{equation}
\theta (- \eta \cdot (x-y))={1\over 2\pi i}\int {dz\over z-i\epsilon}
e^{-iz\eta \cdot (x-y)}
\end{equation}
with $\eta^\mu =(1,0,0,0).$ 

If there are only two external lines, the equation yields the Lehman representation,
and the parameter $z$ can be rewritten as the invariant mass of an 
intermediate state.

For a scattering amplitude $A$, we have of course more than two external lines.  It turns
out that it can be analytically continued $\hat A(z)$ by making some of the momenta
complex through complexifying z and $\eta$.  One can find out the poles and cuts
of $A$ in its kinematical invariants, known as  Mandlestam variables,
by investigating the analyticity of $\hat A(z)$ \cite{book}.  Furthermore, if there are only poles in
$\hat A(z)$ , one will obtain recursion relations, expressing $A$ in terms of lower 
point on-shell scattering amplitudes, as a consequence of Cauchy's theorem.

To give an example, we look at one of the diagrams for  the process 
$P_{1}^{+}P_{2}^{+}P_{3}^{+}
P_{4}^{-}P_{5}^{-}$.  We first write down a largest time equation
\begin{eqnarray}
\Delta (x_{1}-x_{2})\Delta (x_{2}-x_{3})
&=& \big(\theta(-\eta\cdot(x_{1}-x_{3}))\Delta^{+}(x_{1}-x_{2})\nonumber\\
&+&\theta (\eta \cdot (x_{1}-x_{3}))
\Delta^{-}(x_{1}-x_{2})\big)\Delta(x_{2}-x_{3})
\nonumber\\ &+& 
\big(\theta(-\eta\cdot(x_{1}-x_{3}))\Delta^{+}(x_{2}-x_{3})\nonumber\\
&+&\theta (\eta \cdot (x_{1}-x_{3}))
\Delta^{-}(x_{2}-x_{3})\big)\Delta(x_{1}-x_{2}).
\end{eqnarray}
We let $P_{1}$ and $P_{2}$ go into $x_{1}$, $P_{3}$ into $x_{2}$, 
and $P_{4}$ and $P_{5}$ into $x_{3}$, put in the plane wave functions
and integrate over all x's.  Taking out the delta function which enforces 
energy momentum conservation, we have corresponding to  each term
\begin{equation}
{1\over P_{12}^{2}}{1\over P_{45}^{2}}
={1\over P_{12}^{2}}{1\over \hat P_{45}^{2}}\bigg|_{z=z_{12}}
+{1\over \hat P_{12}^{2}}\bigg|_{z=z_{45}}{1\over P_{45}^{2}},
\end{equation}
where 
\begin{eqnarray}
&&\hat P_{1}=P_{1}+z\eta, \ \ \ \hat P_{5}=P_{5}-z\eta,\\
&&\hat P_{12}=\hat P_{1}+P_{2}, \ \ \ \hat P_{45}=P_{4}+\hat P_{5},
\end{eqnarray}
and the on-shell conditions
\begin{eqnarray}
&&\hat P_{12}^{2}=0 \to z_{12}={-P_{12}^{2}\over 2\eta\cdot P_{12}},\\
&&
\hat P_{45}^{2}=0 \to z_{45}={P_{45}^{2}\over 2\eta\cdot P_{45}}.
\end{eqnarray}

Because every term is a rational function in $\eta $, we are allowed 
to maintain this equation when $\eta $ is changed into $N$, the space-cone 
gauge vector.  Let us accept the statement that the vertices
do not change upon the shifts in momenta as described.  We see that
localizing $z$ to different zeros of the kinematical invariants $\hat P_{12}^{2}$
and $\hat P_{45}^{2}$ is to factorize the amplitude into on-shell sub-amplitudes.
\begin{center}
\begin{figure}[!h]
~~~~~~~~~~~~~~~~~~~~~~~\includegraphics[width=4.5in]{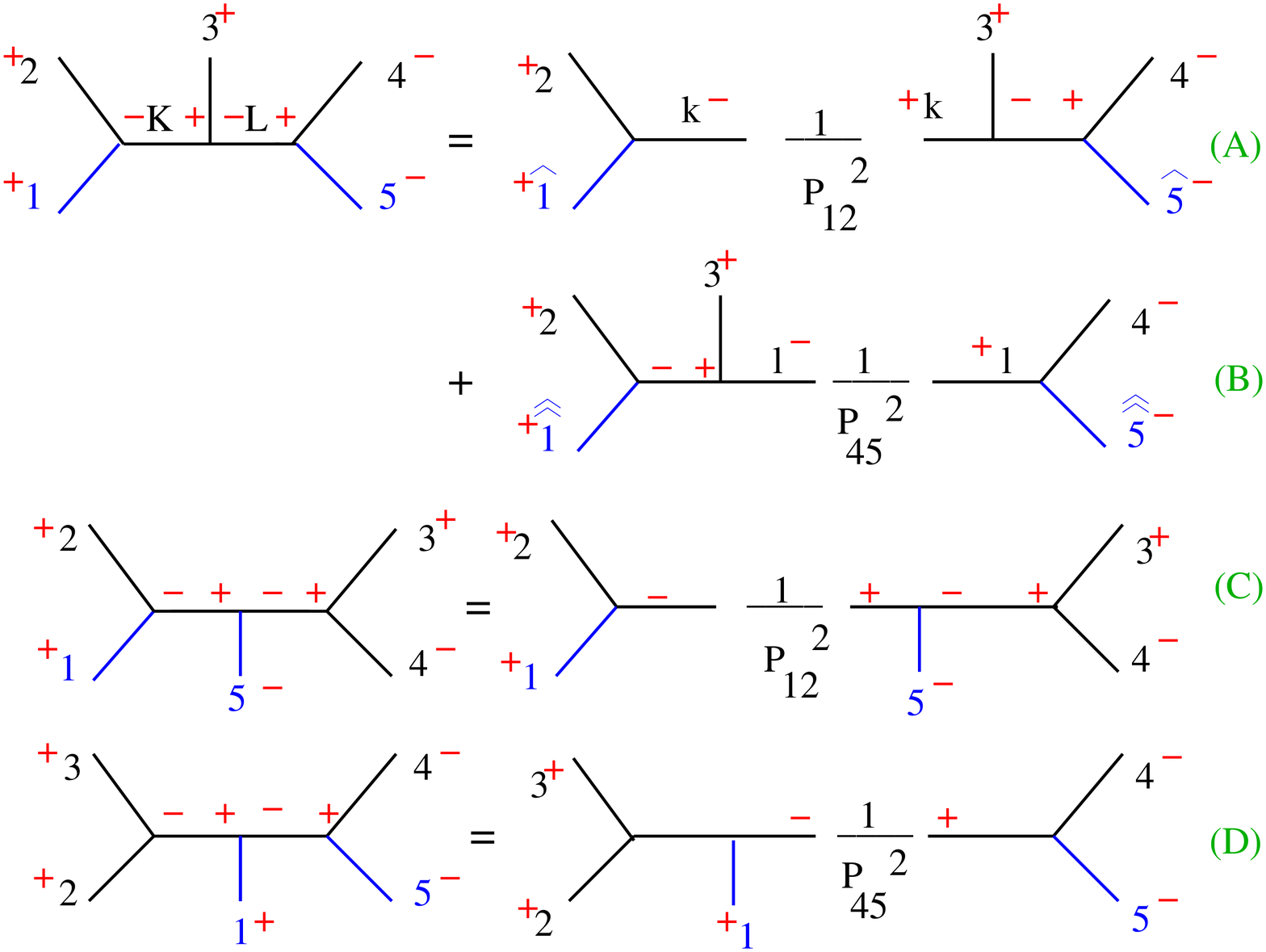}
\caption{Factorization of the 5-point amplitude}
\end{figure}
\end{center}
We have thus a recursion relation, which expresses a five-point amplitude
into a sum of products of three and four-point on-shell amplitudes.

 Furthermore, we see that the above result is also
easily obtained, if we define
\begin{equation}
\hat A(z)={1\over \hat P_{12}^{2}}{1\over \hat P_{45}^{2}},
\end{equation}
and perform the integral 
\begin{equation}
\oint {dz\over z } \hat A(z)=0
\end{equation}
over a  closed contour in the complex a-plane.
The physical amplitude is $\hat A(z=0)$, which is the left hand side,
which is one of the poles in the integral. It is also given
as the sum of the residues due to the other two poles of $\hat A(z).$

\section{Gauge Invariance}

When we use the 'on-shell' method to perform a calculation,
we must be aware that the 'effective vertices' are complexly 
continued lower point amplitudes.  They are made on-shell, 
but they depend on the reference spinors $|+>, [-|.$  The final
result, i. e. the physical amplitudes with real momenta, should 
be independent of any of such choice which is made for 
expediency.  We also recall that fixing $N$ is a choice of gauge
and therefore the independency on $N$ is tantamount to gauge
invariance.

We explore this further in its infinitesimal form.  Suppose 
we first make a choice 
\begin{equation}
|+>, \ \ \ [-|,  \ \ \ \  [-+]=1,
\end{equation}
and then decide to make a small change
\begin{equation}
|+'>=|+>,  \ \  \ |-'|=n([-|+\delta a [x|),
\end{equation}
in which $n$ is a normalization factor and $[x|$ is at this point  
some spinor.  However, when we normalize $[-'+']=1$, we find 
that $[x|=[+|$ is the only solution and therefore
\begin{equation}
[-'|=[-|+\delta  a[+|.
\end{equation}
The requirement of gauge invariance is that physical amplitudes
with real external momenta should be independent of $\delta a$.

We learn from experience that gauge invariance imposes very 
stringent conditions on physical amplitudes.  If there are several
diagrams for a process, gauge invariance relates them in some 
mysterious way to cause tremendous amounts of cancellations
to yield a simple result.  Even though the 'on-shell' method saves
plenty of unnecessary labor, the cancellations are still incomplete.

Let us look at one example.  We analyze a one loop calculation 
of $P_{1}^{-}P_{2}^{+}P_{3}^{+}P_{4}^{+}$, with the choice 
\begin{equation}
|+>=|p_{1}>, \ \ \ [-|=[p_{3}|.
\end{equation}
We would like to make a remark with regard to complexificaton by 
shifting momenta.  In order to reveal all the poles in the kinematical 
invariants that we are interested in, which transcribe into poles in 
the $z$-plane, we must choose shifts and reference spinors properly.
For this example, one set of shifts is
\begin{eqnarray}
&&[\hat 1|=[1|+z[24][-|, \ \ \  |\hat1>=|1>,\\
&&[\hat 2|=[2|, \ \ \ |\hat 2>-|2>+z[4-]||1>,\\
&&[\hat 3|=[3|, \ \ \  |\hat 3>=|3>,\\
&&[\hat 4|=[4|, \ \ \  |\hat 4>=|4>+z[-2]|1>,
\end{eqnarray}
which preserve overall energy momentum conservation.
We find that there are three diagrams which make up this process,
two of which are one particle reducible (1PR) and are easy to 
obtain and one is irreducible and requires some hard calculation.
~~~~~~~~~~~~~~~~~~~~~~~~~~~~~~~~~\begin{center}
\begin{figure}[rhr]
~~~~~~~~~~~~~~~~~~~~~~~~~~~~~~~\includegraphics[width=3.7in]{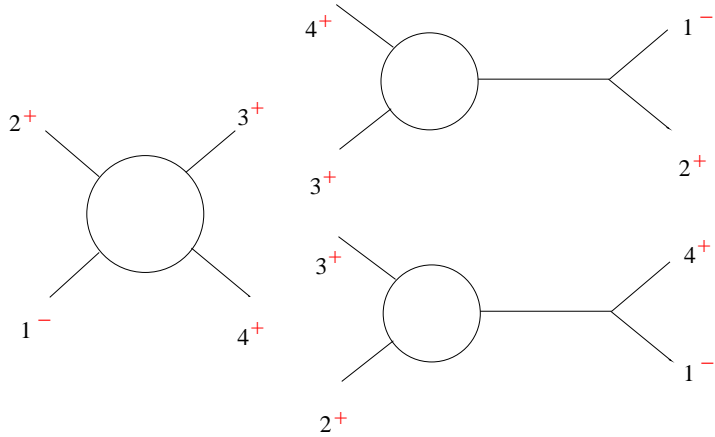}
\caption{The 4-point one-loop (+++-) amplitude}
\end{figure}
\end{center}
From consideration of its collinear  behavior, one can show that
\begin{equation}
A_{4}=A^{1PR}_{4s}f_{s}+A^{1PR}_{4u}f_{u}.
\end{equation}
When we make an infinitesimal gauge change, we have
\begin{equation}
0=\delta A_{4}=[(\delta A^{1PR}_{4s})f_{s}+A^{1PR}_{4s}(\delta f_{s})]
+[(\delta A^{1PR}_{4u})f_{u}+A^{1PR}_{4u}(\delta f_{u})].
\end{equation}
The pole structure in  $s=-(P_{1}+P_{2})^{2}$ and $u=-(P_{1}+P_{4})^{2}$
dictates that each pair of parenthesis should vanish
\begin{equation}
{\delta f_{s} \over f_{s}}=-{\delta A^{1PR}_{4s}\over A^{1PR}_{4s}},
\ \ \    {\delta f_{u} \over f_{u}}=-{\delta A^{1PR}_{4u}\over A^{1PR}_{4u}},
\end{equation}
The right hand sides are known and we can solve these equations to yield
\begin{equation}
f_{s}={-t\over u}, \ \ \ f_{u}={-t\over s}, \ \ \ t=-(s+u).
\end{equation}
We have a recursion relation, which connects four-point ampitudes 
to three-point amplitudes.  $f$'s are known as soft factors, and were postulated in Ref.\cite{Bern:2005hs}.

This line of reasoning can be used for the evaluation of one loop 
amplitudes for $n$ gluons with all but one having the same helicity.
Also, there are recursion relations for the one loop gluon amplitudes with 
all + or all - helicity.  They do not need any soft factors \cite{Bern:2005hs,moriond}.

\section{Concluding Remarks}

There is still much to be uncovered in non-Abelian gauge theories.  We 
have used the freedom of choice in $|+>, [-|$ and complex momentum 
shifts to explore the analyticity of its scattering amplitudes.  The analysis
is further simplified and augmented by the use of space-cone gauge.
Much more can be and needs to be done.

There have been fruitful exchanges and inspiration between non-Abelian 
theories and higher dimensional conformal field theories and strings, 
particularly in calculational aspects of scattering amplitudes 
\cite{Witten:2003nn,Cachazo:2004kj,Britto:2004ap,Maldacena:1997re,Alday:2007hr}.  
Through
these infusions, one may even gain a better understanding of the strong
coupling limit.

We have discussed those amplitudes which are rational functions 
of spinor products.  For more complicated helicity arrangements, there 
is a lot of new developments, such as spinor/twistor integrations,
generalized unitarity, loop integral evaluations, etc.  They can only 
enrich the tool box.

All these are of interest to many of us, because of the relevance
of non-Abelian fields to real physics, a tribute to Professor Yang's
deep insight some fifty years ago.

\end{document}